\PassOptionsToPackage{sort&compress}{natbib}
\documentclass[final,5p,times,twocolumn]{elsarticle}
\usepackage[english]{babel}
\usepackage{amsmath}
\usepackage{amssymb,amsfonts,textcomp}
\usepackage{threeparttable}
\usepackage{tabularx}
\usepackage{longtable}              
\usepackage{multirow,makecell,array}
\usepackage{floatrow}
\usepackage{soul}
\setstcolor{red}
\usepackage{graphicx}
\usepackage[utf8x]{inputenc}
\usepackage{ulem}
\usepackage{hyperref}

\usepackage[version=4]{mhchem}
\graphicspath{{fig/}}
\usepackage{xr}
\begin{document}

\title{SSCHA-based evolutionary crystal structure prediction at finite temperatures with account for quantum nuclear motion}

\author[a_1]{Daniil Poletaev\corref{cor1}}\ead{d.poletaev@skoltech.ru}\ead{poletaev.dan@gmail.com}
\author[a_1,a_2]{Artem R. Oganov}%
\address[a_1]{Materials Discovery Laboratory, Skolkovo Institute of Science and Technology, 121205 Moscow, Russian Federation} 
\address[a_2]{SberUniversity, 143581 Anosino, Istra Urban District, Russian Federation} 

\cortext[cor1]{Corresponding author: Daniil Poletaev.}

\date{\today}

\begin{abstract}

Reliable crystal structure prediction (CSP) at finite temperatures, including quantum anharmonic effects, remains challenging but is particularly important for systems containing light atoms such as superconducting hydrides. Here, we integrate machine-learned interatomic potentials (MLIPs) with the stochastic self-consistent harmonic approximation (SSCHA) to enable evolutionary CSP on the quantum anharmonic free-energy landscape. Using LaH$_{10}$ as a test case, we compare three approaches: (i) active-learning MLIPs (AL-MLIPs) trained on the fly, (ii) universal MLIPs (uMLIPs), and (iii) temperature-dependent effective potentials (TDEPs) trained on SSCHA ensemble data. AL-MLIPs correctly predict the cubic Fm$\bar{3}$m phase but require thermodynamic perturbation theory corrections for consistent results. The foundation uMLIP Mattersim-5m enables SSCHA-based CSP without per-structure training, but fine-tuning is required for higher accuracy. Temperature-dependent effective potentials trained on SSCHA data enable accurate and very efficient CSP for large unit cells on the quantum anharmonic free-energy landscape. We demonstrate that quantum anharmonicity simplifies the free-energy landscape and is essential for the correct energy ranking of LaH$_{10}$ structures.

\end{abstract}

\begin{keyword}

crystal structure prediction \sep
structure ranking at finite temperature \sep
quantum anharmonicity \sep
La-H \sep
thermodynamic perturbation theory \sep 
USPEX \sep 
moment tensor potentials \sep
active learning \sep
universal MLIPs \sep
superconducting hydrides \sep
effective potentials

\end{keyword}

\maketitle

\section{Introduction}
\label{sec:introduction}

Accurate prediction of crystal structures at various pressures and temperatures is crucial for materials discovery and design \cite{Oganov2006, Oganov2010}. The success of structure prediction depends critically on correctly solving both the structure sampling and the structure ranking problems \cite{Oganov2010}. While state-of-the-art \textit{ab initio} methods, e.g. density functional theory (DFT) and post-DFT methods provide the highest accuracy for structure ranking at 0 K, their computational cost severely limits applications at finite temperatures \cite{Oganov2006, Oganov2010, Hunnisett2024, Wang2010, VanderVen1998, Fiedler2023}.
Most computational searches therefore rely on structure ranking exclusively at 0 K \cite{Hunnisett2024, Kruglov2023}.

T-USPEX was the first developed technique capable of overcoming these limitations by enabling finite-temperature structure prediction using classical molecular dynamics (MD) with semi-empirical or machine-learning interatomic potentials (MLIPs) for structure relaxation, sampling phase space, and free energy calculation combined with pressure and free energy corrections within the thermodynamic perturbation theory for ensuring structure ranking at \textit{ab initio} level \cite{Kruglov2023}. However, classical MD is unsuitable for materials where quantum nuclear motion is essential. While harmonic systems can be treated by lattice dynamics methods, quantum anharmonic systems remain challenging. Many systems, particularly superconducting hydrides \cite{Errea2014pss, Errea2014prb, Errea2020, Meninno2022}, charge-density-wave materials \cite{Bianco2017, Rivano2025arxiv}, and thermoelectric compounds \cite{Monacelli2021}, exhibit quantum anharmonicity that critically influences thermodynamic stability and material properties both at 0~K and finite temperatures.

The stochastic self-consistent harmonic approximation (SSCHA) provides a rigorous framework for treating quantum anharmonic effects \cite{Errea2014prb, Monacelli2018, Monacelli2021}.
However, the computational cost of SSCHA prohibits its direct use in structure prediction. 
DFT-based structure relaxation within SSCHA can take weeks to months, whereas hundreds to thousands of such optimizations are usually required \cite{Oganov2010}.

Active-learning MLIPs (AL-MLIPs) offer a promising solution for accelerating SSCHA calculations \cite{Lucrezi2023, Belli2025, Rivano2025arxiv, Schutt2018, Ranalli2023, Zhao2025arxiv}. 
When integrated with structure prediction, AL-MLIPs can dramatically reduce the cost of SSCHA relaxations while preserving the accuracy needed to model quantum anharmonicity at \textit{ab initio}-level.
Moreover, the recent development of universal MLIPs (uMLIPs), also known as large atomistic models or foundation models \cite{Riebesell2025}, presents a further opportunity to make crystal structure prediction (CSP) with quantum anharmonic effects more feasible.
Finally, the effective potentials or free energy models emerge as a new promising way to perform high-throughput CSP on quantum anharmonic free energy landscapes at finite temperatures \cite{Wang2026}.

In this work, we integrate AL-MLIPs, uMLIPs, and effective potentials into the SSCHA-based structure prediction within USPEX \cite{Oganov2006, Oganov2010}. We explore three distinct strategies: (i) active-learning MLIPs trained locally per structure or reused across structures, (ii) foundation uMLIPs applied without per-structure training, and (iii) temperature-dependent effective potentials (TDEPs) trained on SSCHA ensemble data. 
% to enable scalable CSP with large unit cells.
Using LaH$_{10}$ at 150~GPa as a test case, where phase stability is governed by quantum anharmonicity, we compare evolutionary CSP on the quantum anharmonic free energy landscape using these approaches and demonstrate that employing the effective potentials is the most efficient way, not only providing a significant speedup but also successfully overcoming the unit-cell-size scalability problem.

LaH$_{10}$ has been extensively studied both theoretically and experimentally \cite{Liu2017a,Peng2017,Liu2018a,Kruglov2020,Errea2020,Ly2022,Bund2023,Guo2024,Geballe2017,Somayazulu2019,Drozdov2019,Sun2021a,Laniel2022}, enabling robust validation of our CSP results. The cubic $\mathrm{Fm}\bar{3}\mathrm{m}$ phase of LaH$_{10}$ with sodalite-like structure was first theoretically predicted at pressures $\geq 150$ GPa \cite{Liu2017a,Peng2017} and was found experimentally to be stable at pressures above 135--140 GPa \cite{Geballe2017,Sun2021a,Laniel2022}. Theoretical studies at different levels of theory have revealed that accounting for anharmonic effects is essential: while the cubic phase appears dynamically unstable within the harmonic approximation at pressures below 210--230 GPa, anharmonic effects stabilize it at pressures of 129--150 GPa \cite{Liu2018a,Errea2020,Kruglov2020}.

\section{Results}
\label{sec:results}

We performed two independent searches for the most stable structure of LaH$_{10}$ on the quantum anharmonic free energy landscape at 150~GPa and 300~K using AL-MLIPs, one search using the best-performing uMLIP, and three searches with effective potentials trained on the data retained from the CSP with the uMLIP.  

In the CSPs with AL-MLIPs and uMLIP, for each structure proposed by USPEX, we performed a SSCHA relaxation by minimizing the free-energy functional $F[\mathcal{R},\Phi]$ with respect to the centroid positions $\mathcal{R}$ and the auxiliary force-constant matrix $\Phi$ using MLIP for calculation of energies, forces, and stresses in statistical ensembles:
\begin{equation}
F = \min_{\mathcal{R},\,\Phi} \mathcal{F}[\mathcal{R}, \Phi]
  = \mathcal{F}[\mathcal{R}_{\mathrm{eq}}, \Phi_{\mathrm{eq}}].
\end{equation}

\begin{figure}
    \centering
    \includegraphics[width=1\linewidth]{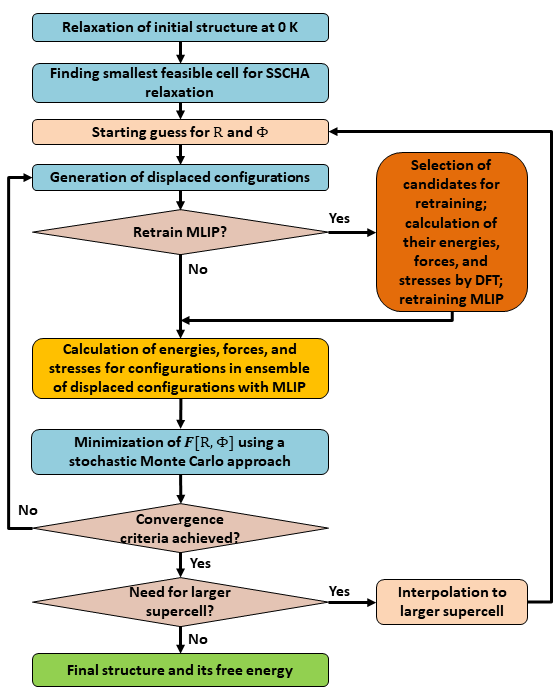}
    \caption{A common scheme for structure optimization within the SSCHA using an active-learning MLIP.}
    \label{fig:sscha_mlip_common_scheme}
\end{figure}

A schematic of the SSCHA relaxation workflow using retrainable MLIP is shown in Figure~\ref{fig:sscha_mlip_common_scheme}. 
To reduce the computational cost of SSCHA, we used a multistage upscaling scheme similar to those described in the works of Lucrezi et al.~\cite{Lucrezi2023} and Beli and Zurek \cite{Belli2025}. 
The details of the SSCHA relaxation using this scheme are described in the "Methods" section. 

\subsection{Active-learning MLIPs for SSCHA-based CSP}
\label{sec:mlip_strategies_proposed}

We explored three strategies for applying active learning to MLIPs during SSCHA relaxations within the CSP workflow, when several structures may be treated in parallel and several MLIPs may therefore be trained simultaneously. 

\textbf{Strategy 1 (Local per-structure training):} For each new structure, a separate MLIP is trained from scratch on configurations from its SSCHA ensembles (see Figure S1 in the Supplementary materials). While computationally feasible and allowing high-accuracy MLIPs tailored to each structure, this approach risks redundant DFT calculations for structures that existing MLIPs could already describe adequately.

\textbf{Strategy 2 (Global MLIP retraining):} A single global MLIP is retrained for each structure during a generation, then all retraining configurations are aggregated and the MLIP is retrained again once per generation (see Figure S2 in the Supplementary materials). This minimizes redundant DFT calculations but has a critical drawback: retraining global MLIPs (for example, high-level moment-tensor potentials (MTPs) \cite{Shapeev2016,Podryabinkin2023}, see Figure S3) can demand more resources than the DFT calculations themselves. 

\textbf{Strategy 3 (Hybrid: Reuse and retrain from existing local MLIPs):} For the first batch of structures treated in parallel, a separate MLIP is trained from scratch for each structure (see Figure S4 in the Supplementary Materials). All trained MLIPs are then stored. For subsequent structures, a selection algorithm (MaxVol \cite{Olshevsky2010} for MTPs) identifies the stored MLIP that requires the fewest additional configurations for retraining. This MLIP is then selected and retrained for the new structure. This strategy reduces the number of DFT calculations needed to train MLIPs for new structures without substantially increasing the computational cost of the training procedure, because lightweight MLIPs can be used. The drawback is that the number of local MLIPs grows with each generation; however, fast selection algorithms mitigate this cost.

The explored strategies, their advantages and drawbacks are described in more details in the Supplementary Materials.
In this work, we consider the results of CSP obtained with Strategy 1 (employing local AL-MLIPs) for SSCHA-based structure prediction as a reference, compare them with the results obtained with Strategy 3 (employing reusable AL-MLIPs), and completely discard Strategy 2 (employing global AL-MLIP) due to its inefficiency.

\subsection{CSP with local AL-MLIPs}
\label{sec:results_local_mlips}

\begin{figure}
    \centering
    \includegraphics[width=1\linewidth]{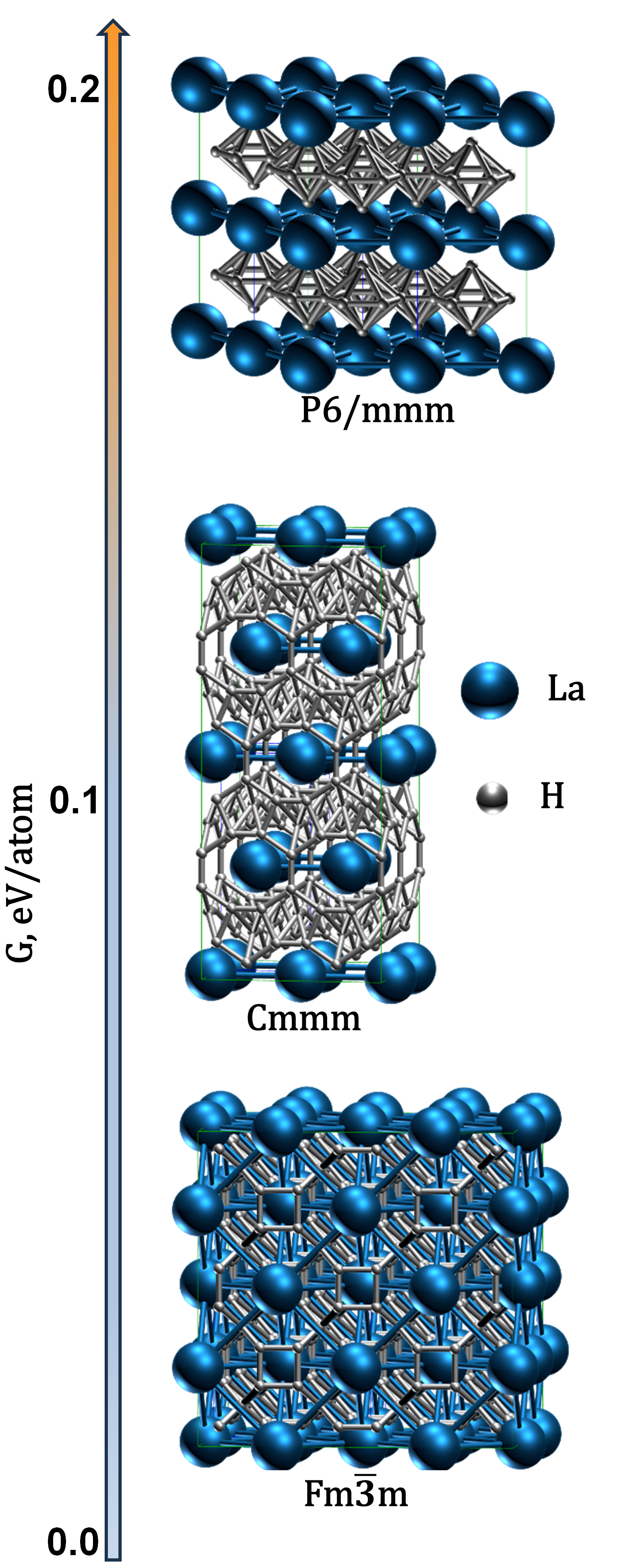}
    \caption{Three LaH$_{10}$ polymorphs with $\mathrm{Fm}\bar{3}\mathrm{m}$, $\mathrm{Cmmm}$, and $\mathrm{P}6/\mathrm{mmm}$ space groups predicted on the anharmonic free energy landscape at 300~K and 150~GPa with AL-MLIPs trained from scratch for each structure. 
    The values of the free energy, $G$, are with respect to the ground state structure, $\mathrm{Fm}\bar{3}\mathrm{m}$.
    The crystal structures were visualized using the STMng program \cite{Valle2005}.
    } 
    \label{fig:fig_lah10_anharmonic_structures}
\end{figure}

With local AL-MLIPs trained from scratch for each structure, USPEX predicted only three unique phases of LaH$_{10}$ on the anharmonic free energy landscape. 
The crystallographic information for these phases and visualizations of their structures are presented in Table \ref{tab:lah10_phases_info} and Figure \ref{fig:fig_lah10_anharmonic_structures}.

\begin{table}[h]
\caption{Crystallographic information about three LaH$_{10}$ phases predicted on the anharmonic free energy landscape at 150~GPa and 300~K with AL-MLIPs trained from scratch for each structure. Fitness is a value of the free energy with respect to the ground state at the given conditions.}
\label{tab:lah10_phases_info}
\footnotesize
\begin{tabularx}{\columnwidth}{lccllll}
\hline
Phase & Fitness &  Lattice  & \multicolumn{4}{c}{Atom coordinates} \\
 &  (eV/atom) & parameters &   \\
\hline
$Fm\bar{3}m$ & 0 & $a = 5.120$ \AA & La & 0.0 & 0.0 & 0.0 \\
            &  &                 & H & 0.25 & 0.25 & 0.25 \\
            &  &                 & H & 0.38 & 0.38 & 0.38 \\
\hline
$Cmmm$ & 0.085  & $a = 2.932$ \AA   & La & 0.0 & 0.0 & 0.0 \\
       & & $b = 7.020$ \AA   & H & 0.0 & 0.295 & 0.153 \\
       & & $c = 3.276$ \AA   & H & 0.0 & 0.421 & 0.5 \\
       & &                  & H & 0.168 & 0.172 & 0.5 \\
\hline
$P6/mmm$ & 0.160 & $a = 3.248$ \AA & La & 0.0 & 0.0 & 0.0 \\
         & & $c = 3.735$ \AA & H & 0.333 & 0.667 & 0.239 \\
         & &               & H & 0.314 & 0.0 & 0.5 \\
\hline

\end{tabularx}
\end{table}

The most stable predicted polymorph has cubic $\mathrm{Fm}\bar{3}\mathrm{m}$ symmetry, in agreement with the experimental data \cite{Drozdov2019}, and already appeared in the first generation.
It should be noted that, \textit{before} SSCHA relaxation, this polymorph had lower symmetry (Immm), i.e., in the structure-prediction results on the classical energy landscape at 0~K it would appear as an Immm structure rather than $\mathrm{Fm}\bar{3}\mathrm{m}$.
In subsequent generations, cubic $\mathrm{Fm}\bar{3}\mathrm{m}$ LaH$_{10}$ was obtained several times after SSCHA relaxation of structures with P$\bar{1}$, C$2$, C$2$/m, Immm, I$4$/mmm, R$\bar{3}$m, and $\mathrm{Fm}\bar{3}\mathrm{m}$ symmetries on the classical energy landscape; see Table~S1 in the Supplementary Materials for details.
The calculated lattice parameter for cubic LaH$_{10}$ (5.120~$\AA$) shows a slight deviation from the experimental value (5.102~$\AA$) measured at 150~GPa \cite{Drozdov2019}.

The second LaH$_{10}$ structure predicted on the anharmonic free energy landscape has orthorhombic $\mathrm{Cmmm}$ symmetry and a Gibbs free energy at 300~K that is about 86~meV/atom higher than that of the cubic $\mathrm{Fm}\bar{3}\mathrm{m}$ structure.
It was obtained from structures with P$1$ and $\mathrm{Cmmm}$ space groups that emerged on the classical energy landscape.
The third predicted LaH$_{10}$ polymorph has a layered hexagonal structure with $\mathrm{P}6/\mathrm{mmm}$ symmetry.
It was relaxed within the SSCHA from the $\mathrm{P}6/\mathrm{mm}$ structure and lies about 160~meV/atom above the ground state.

\begin{figure}
    \centering
    \includegraphics[width=1\linewidth]{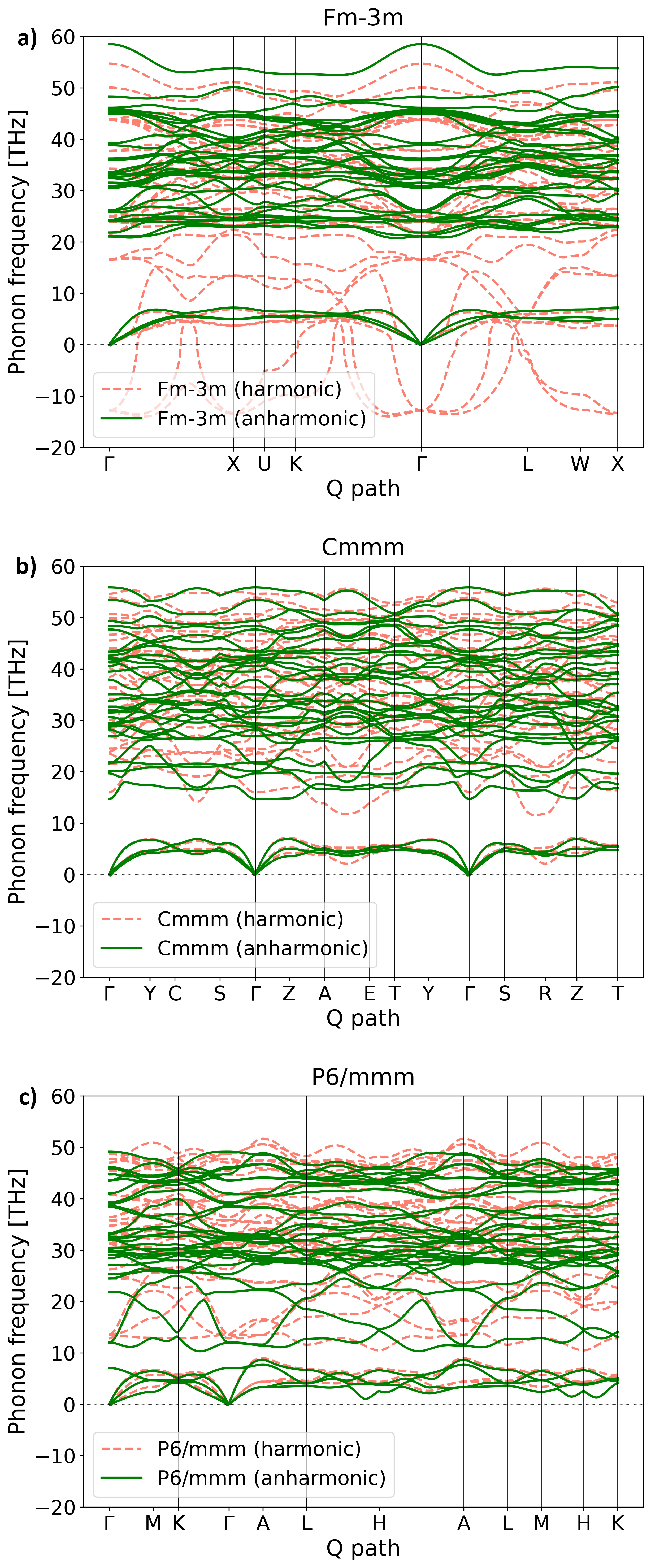}
    \caption{Harmonic and anharmonic phonon dispersion curves at 150~GPa for: a) cubic LaH$_{10}$ with $\mathrm{Fm}\bar{3}\mathrm{m}$ symmetry group; b) orthorhombic LaH$_{10}$ with $\mathrm{Cmmm}$ symmetry group; and c) hexagonal LaH$_{10}$ with $\mathrm{P}6/\mathrm{mmm}$ symmetry group.
    } 
    \label{fig:fig_lah10_cmmm_harmonic_anharmonic_phonon_dispersion}
\end{figure}

The phonon dispersions clearly show that the lowest-energy cubic $\mathrm{Fm}\bar{3}\mathrm{m}$ polymorph is dynamically stabilized by anharmonic lattice vibrations, consistent with previous literature \cite{Errea2020,Laniel2022} (see Figure \ref{fig:fig_lah10_cmmm_harmonic_anharmonic_phonon_dispersion}). In contrast, the higher-energy orthorhombic $\mathrm{Cmmm}$ and hexagonal $\mathrm{P}6/\mathrm{mmm}$ polymorphs are dynamically stable both harmonically and anharmonically.

%%%%%%%%%%%%%%%%%%%%%
\subsection{CSP with reusable AL-MLIPs}
\label{sec:results_reusable_mlips}

CSP for the most stable LaH$_{10}$ polymorph employing reusable AL-MLIPs yielded results identical to those obtained with local AL-MLIPs trained from scratch within the 100~meV/atom energy threshold above the ground state, i.e., predicting the $\mathrm{Fm\bar{3}m}$ and $\mathrm{Cmmm}$ structures.
However, beyond this threshold, the predicted structures differed.
The hexagonal $\mathrm{P}6/\mathrm{mmm}$ structure, with a Gibbs free energy 160~meV/atom above the ground state, was not predicted with reusable AL-MLIPs. Instead, three other structures with $\mathrm{C}2/\mathrm{m}$, $\mathrm{R}3\mathrm{m}$, and $\mathrm{Cmmm}$ space groups, with Gibbs free energies 126, 162, and 186~meV/atom above the ground state, were predicted; see Table~S2.
This is expected, as evolutionary algorithms narrow the search space with each generation, focusing search on low-energy structures. This underscores the importance of structure prediction on finite-temperature free-energy landscapes, as this is the only way to identify structures that are far from the 0~K ground state but may be stabilized at higher temperatures.

We, however, encountered caveats when employing reusable AL‑MLIPs, which highlight two important considerations.
First, repeatedly retraining a lightweight AL‑MLIP (e.g., a level-eight MTP, as in our case) on a continuously updated dataset can lead to accumulated training errors.
These errors degrade the MLIP's predictions of energies, forces, and stresses for perturbed configurations generated during SSCHA relaxations. 
In particular, we noticed that SSCHA relaxation employing reusable MTPs with large errors in energies, forces, and stresses could converge even when the structure was not fully relaxed, unlike relaxation performed with a more accurate MTP.
For example, in our case this led to the prediction of structures with the $\mathrm{Fm\bar{3}m}$ space group that had the same topology as the ground-state $\mathrm{Fm\bar{3}m}$ structure. However, their atomic volumes were smaller by $0.04-0.20$~\AA$^{3}$/atom, and their thermodynamically corrected free energies were higher by $8-37$~meV/atom than those of the ground-state structure.
In Table S2, we have marked such structures in red. 
Further SSCHA relaxations of these structures (started from their final auxiliary dynamical matrices) using more accurate local AL-MLIPs trained from scratch converged only when their atomic volumes and thermodynamically corrected free energies matched those of the ground state within confidence intervals of 0.02~\AA$^{3}$/atom and 2~meV/atom (see Table~S3).

Additionally, we investigated whether the $\mathrm{C}2/\mathrm{m}$, $\mathrm{R}3\mathrm{m}$, and $\mathrm{Cmmm}$ structures with free energies exceeding 100~meV/atom above the ground state can be relaxed further with more accurate MLIPs, given that the MTPs used for their SSCHA relaxation also had large errors. 
Indeed, after relaxation of these structures with more accurate local AL-MLIPs, their free energies decreased and their atomic volumes also changed, similar to the case of the $\mathrm{Fm\bar{3}m}$ structures.
Interestingly, when SSCHA relaxation of the $\mathrm{C}2/\mathrm{m}$ structure with local AL-MLIPs trained from scratch was continued from the state obtained with the crude AL-MLIP, the structure relaxed into a deeper local minimum on the quantum anharmonic free-energy landscape, but its space group did not change. In contrast, when SSCHA relaxation of this structure with a more accurate MLIP started directly from the 0~K local minimum on the Born-Oppenheimer energy landscape, it transformed into the ground-state cubic $\mathrm{Fm\bar{3}m}$ phase.
This further explains why this $\mathrm{C}2/\mathrm{m}$ structure was not predicted in the CSP with local AL-MLIPs trained from scratch and highlights the importance of using an accurate MLIP(s) from the outset if one aims to relax structures generated by a CSP algorithm to the deepest possible minima.

The second consideration is that without corrections from thermodynamic perturbation theory~\cite{Landau2013,Vocadlo2002,Vocadlo2004,Oganov2004,Kruglov2023}, the maximal dispersion in the calculated free energies for ostensibly identical structures relaxed with reusable AL-MLIPs can exceed 100~meV/atom. 
In our case, it reached 161~meV/atom for the lowest-energy $\mathrm{Fm\bar{3}m}$ structures, even without taking into account the structures with large differences in atomic volume.
Applying these corrections reduced the maximal dispersion to 7~meV/atom. 
For comparison, with local AL-MLIPs trained from scratch for each structure, the maximal dispersion of the free energy for the $\mathrm{Fm\bar{3}m}$ structures was 22~meV/atom without corrections and was reduced to 2~meV/atom after corrections were applied. 
For more detailed information about the corrections applied to each structure predicted with local and reusable MTPs, we refer the reader to Tables~S1 and S2.

Thus, applying pressure and free-energy corrections within thermodynamic perturbation theory is essential when employing lightweight reusable AL-MLIPs, which tend to accumulate generalization errors.
Even then, the accuracy of CSP with reusable AL-MLIPs can be unsatisfactory.
A dispersion of 7~meV/atom is acceptable when the free-energy differences between unique predicted structures on a quantum anharmonic landscape are on the order of tens of meV/atom, as in LaH$_{10}$, which is a relatively simple system.
However, for more complex systems in which the free-energy differences between polymorphs can be only a few meV/atom, such accuracy is clearly insufficient, and the use of local AL-MLIPs or more accurate reusable AL-MLIPs (e.g., a level-twenty MTP instead of a level-eight MTP) is better justified despite the higher computational cost.

The main advantage of reusing AL-MLIPs is that it reduces the number of DFT calculations required for retraining.
The ratio of total MLIP evaluations (energies, forces, and stresses) during SSCHA relaxations to the total number of DFT calculations required for retraining exceeded 2000 for reusable AL-MLIPs, compared with only $\sim$600 for MLIPs trained from scratch. This corresponds to a more than threefold reduction in the number of required DFT calculations.
Overall, using AL-MLIPs in SSCHA relaxations reduces the number of DFT calculations by two (in the case of local AL-MLIPs) to three (in the case of reusable AL-MLIPs) orders of magnitude.

\subsection{CSP with universal MLIPs}
\label{sec:results_umlips}

The proposed strategies for combining AL-MLIPs with SSCHA-based structure prediction were developed for cases where suitable interatomic potentials are not available or very crude for the desired accuracy. Until recently, this problem was acute; however, now a variety of so-called universal machine-learned interatomic potentials (uMLIPs) demonstrate significant improvements in describing materials properties at the atomic level \cite{Riebesell2025}. 
In this work, we checked if any of the existing uMLIPs can be used in the high-pressure CSP on anharmonic free energy landscape out-of-the-box.
We considered several uMLIPs from the Matbench discovery project \cite{Riebesell2025}, namely CHGNet \cite{Deng2023}, DPA3 \cite{Zhang2025}, GRACE \cite{Bochkarev2024}, MACE \cite{Batatia2024}, Mattersim-5m \cite{Yang2024}, OrbNet-v2 \cite{Neumann2024}, and Sevennet \cite{Kim2024}.

Before conducting finite-temperature CSP with a uMLIP, we qualitatively evaluated each potential by relaxing the rhombohedral ($\mathrm{R\bar{3}m}$) LaH$_{10}$ structure within the SSCHA at 150~GPa and 300~K.
With a sufficiently good potential, it should transform into the cubic ($\mathrm{Fm\bar{3}m}$) structure.
Only three uMLIPs, namely CHGNet \cite{Deng2023}, MACE \cite{Batatia2024}, and Mattersim-5m \cite{Yang2024}, successfully completed the SSCHA relaxation at the given conditions without failing due to unphysical atomic configurations. 
Among these, only Mattersim-5m relaxed the structure into the experimentally known stable cubic phase ($\mathrm{Fm\bar{3}m}$).
Based on these results, we selected Mattersim-5m for further testing in CSP on the anharmonic free-energy landscape of LaH$_{10}$ at 150~GPa and 300~K using USPEX.

In the subsequent SSCHA-based USPEX search powered by Mattersim-5m uMLIP, the cubic $\mathrm{Fm\bar{3}m}$ structure emerged in the first generation, consistent with previous searches using AL-MLIPs. It appeared multiple times, evolving from initial structures with lower ($\mathrm{Fmmm}$, $\mathrm{Immm}$, $\mathrm{I\bar{4}m2}$, $\mathrm{Cm}$) or the same ($\mathrm{Fm\bar{3}m}$) symmetry; see Table S6 for details. 
The free energy of this cubic phase was the lowest among all converged structures.

The second-ranked structure identified by Mattersim-5m on the anharmonic landscape has $\mathrm{Cmmm}$ symmetry, and it is almost identical to the one found with AL-MLIPs though having different lattice parameters.
Without thermodynamic perturbation theory corrections, Mattersim-5m placed this structure closer to the ground state (26~meV/atom above it, compared with 86~meV/atom with AL-MLIPs), but after applying these corrections its free energy became 104~meV/atom higher than that of the ground state.

Twelve additional structures with symmetries $\mathrm{C2/m}$, $\mathrm{Immm}$, $\mathrm{Amm2}$, $\mathrm{R3m}$, $\mathrm{P2/m}$, $\mathrm{C222}$, and $\mathrm{P4/mmm}$, all with Gibbs free energies more than 100~meV/atom above the ground state after applying thermodynamic perturbation theory corrections, were also found on the anharmonic free-energy landscape with Mattersim-5m.
Most of these structures (namely, those with $\mathrm{Immm}$, $\mathrm{Amm2}$, $\mathrm{P2/m}$, $\mathrm{C222}$, and $\mathrm{P4/mmm}$ symmetries) were not found after the CSP with local and reusable AL-MLIPs. 
Initially, we suggested that they could represent false local minima stemming from insufficient accuracy of the MLIPs, not from peculiarities of the quantum anharmonic free energy landscape, because errors in energies, forces, and stresses of the foundation model Mattersim-5m, evaluated on the configurations used to calculate thermodynamic perturbation theory corrections for these structures, are comparable to those of the least accurate reusable AL-MLIPs; see Table~S4 for details.
If this is true, they should relax further into the true minima on the quantum anharmonic free energy landscape with a more accurate MLIP.

To investigate how using a more accurate MLIP will affect relaxation of these metastable structures, we fine-tuned the foundation model Mattersim-5m on the dataset obtained from the DFT calculations used to evaluate thermodynamic perturbation theory corrections for these structures. We then continued their SSCHA relaxation using the fine-tuned model, starting from the final auxiliary dynamical matrices obtained with the original Mattersim-5m model.
After fine-tuning Mattersim-5m, the MLIP errors for all these structures became much smaller, as did the pressure and free-energy corrections.
In particular, the free energy and volume of the second-ranked $\mathrm{Cmmm}$ structure became almost identical to those obtained with a local AL-MLIP.
Three structures changed their symmetries ($\mathrm{C222}\to\mathrm{C2/m}$, $\mathrm{C2/m}\to\mathrm{Immm}$, and $\mathrm{P2/m}\to\mathrm{Imm2}$) while other structures only changed their volumes and free energies.
Some structures changed their ranking after additional SSCHA relaxation with fine-tuned Mattersim-5m.
Here, we emphasize that the initial structure ranking obtained after SSCHA relaxations with the foundation Mattersim-5m model and free-energy corrections within thermodynamic perturbation theory was also correct.
The ranking changed after relaxation with fine-tuned Mattersim-5m because deeper local minima were found.
For more detailed information, we refer the reader to Table~S5.
What is more important, none of these twelve new structures were relaxed into the minima previously found within CSP using AL-MLIPs, even with fine-tuned Mattersim-5m, indicating that with Mattersim-5m one can find a larger number of diverse polymorphs on the quantum anharmonic free energy landscape than with lightweight AL-MLIPs.

Overall, we found that the foundation uMLIP Mattersim‑5m can robustly guide SSCHA-based crystal structure prediction (CSP), even at high pressures, while, with the help of thermodynamic perturbation theory, providing \textit{ab initio} structure ranking and nearly eliminating the need to train AL‑MLIPs from scratch. However, fine-tuning of this uMLIP is still required to obtain structure-relaxation results closer to the DFT level of theory, because thermodynamic perturbation corrections do not resolve inaccuracies in the SSCHA relaxation itself.

The advantage of Mattersim‑5m over lightweight AL‑MLIPs is that, with this uMLIP, roughly 90\% of structures pre-relaxed at 0~K on the classical energy landscape also relax to a converged state within SSCHA at 300~K. In contrast, level‑eight MTPs used as AL‑MLIPs lead to convergence in SSCHA relaxation for only about 20--30\% of such structures.
Another benefit is that, when a single uMLIP is used, thermodynamic perturbation corrections are needed only to ensure \textit{ab initio} structure ranking; unlike in the case of local and reusable lightweight AL-MLIPs, they are not essential for consistency of free energies and volumes of duplicate structures. 
This makes it possible to apply thermodynamic corrections only to the unique structures found on the anharmonic free-energy landscape, as was done in this work, thereby reducing the number of DFT calculations by an order of magnitude comparing to the case of AL-MLIPs.

\subsection{Effective potentials for scalable CSP}
\label{sec:results_effective_mlips}

The searches with AL-MLIPs and uMLIP described above used an 11-atom unit cell, which is sufficient for LaH$_{10}$. However, in real CSP for unexplored systems, the optimal unit-cell size is unknown, and larger cells must be explored. Direct SSCHA relaxation of large cells (e.g., 88 atoms) is prohibitively expensive even with MLIP acceleration. 

To overcome this, we trained temperature-dependent effective potentials (TDEPs) from SSCHA ensemble data. After each SSCHA relaxation step, the weighted-averaged forces, stresses, and free energy can be saved as a training point for a potential that effectively describes the anharmonic free-energy landscape. With sufficient data, such a potential allows structure optimization at finite temperature as easily as a classical potential at 0~K. A related approach was recently proposed by Wang et al. \cite{Wang2026} as deep free energy (DF) models.

We collected weighted-averaged data from SSCHA ensembles obtained with the Mattersim-5m uMLIP, and trained TDEPs within both the MTP \cite{Shapeev2016} and DPA-3 \cite{Zhang2025} frameworks. MTP-based TDEPs, even at highest levels, exhibited training RMSE of $\sim$65~meV/atom and poor transferability to larger cells, resulting in incorrect structure ranking after CSP with an 11-atom cell and failed CSP with 44- and 88-atom cells. This highlights that the MTP framework, while effective for classical AL-MLIPs, is not well-suited for TDEPs in this context.

In contrast, TDEP-DPA3 models achieved excellent accuracy (RMSE 0.8–3.8~meV/atom) and successfully guided CSP with 11-, 44-, and 88-atom unit cells, always predicting the experimentally observed cubic $\mathrm{Fm\bar{3}m}$ structure as the most stable. The deep free energy (DF) model for La-Sc-H system from \cite{Wang2026} also succeeded, finding the cubic structure even faster (1st, 1st, and 5th generations for 11, 44, and 88 atoms, respectively, versus 2nd, 11th, and 18th for our TDEP-DPA3). The better performance of the DF model can be attributed to its wider training domain and the use of a more accurate Born-Oppenheimer MLIP for computing energies, forces, and stresses in SSCHA ensembles.

For comparison, classical 0~K CSP with 11-, 44-, and 88-atom cells using VASP for structural relaxation resulted in finding low-symmetry structures as the most stable instead of the cubic $\mathrm{Fm\bar{3}m}$ structure. This confirms that quantum anharmonicity is essential for correct structure ranking, and that effective potentials trained on SSCHA data provide a scalable pathway to incorporate these effects in CSP.

The effective potential approach thus directly addresses the unit-cell-size scalability problem, making it possible to perform CSP on quantum anharmonic free-energy landscapes with large unit cells at a fraction of the cost of direct SSCHA relaxations.

\section{Discussion}
\label{sec:discussion}

We carried out searches for the most stable LaH$_{10}$ polymorph on the quantum anharmonic free-energy landscape using three distinct strategies: local and reusable AL-MLIPs, the foundation uMLIP Mattersim-5m, and temperature-dependent effective potentials (TDEPs) trained on SSCHA ensemble data. The first two strategies were applied with 11-atom unit cells; the third successfully applied also with 44- and 88-atom cells.

A CSP with direct SSCHA relaxations for each structure in the first two strategies, while accurate due to the thermodynamic perturbation theory corrections, is computationally demanding: roughly two orders of magnitude more expensive than a DFT-based 0~K CSP, even with a ready-to-use uMLIP. 
The TDEP approach presented in this work provides a scalable pathway for finite-temperature CSP. By training effective potentials on weighted-averaged SSCHA data, we bypass the expensive direct SSCHA relaxation of each structure during CSP. Using TDEP-DPA3, we successfully performed CSP with up to 88 atoms per cell, correctly identifying the cubic $\mathrm{Fm}\bar{3}\mathrm{m}$ structure as the ground state. This capability is essential for real-world CSP where the optimal unit-cell size is unknown.

The bottleneck of the TDEP approach is the need to generate data and train TDEPs for each temperature separately, which could be a time-consuming task.
In this work we used the data that was generated after a sufficiently long CSP within the second strategy (USPEX+SSCHA+MatterSim-5m), but for practical CSP cases more efficient approaches to quantify the free energy landscapes, allowing to minimize the number of costly SSCHA relaxations, should be developed.
Even though such approaches are beginning to emerge~\cite{Wang2026} the overall cost of training TDEP and performing CSP with TDEP would yet be greater than performing 0~K CSP.

A cheaper way for exploring quantum anharmonic free-energy landscapes could be to perform 0~K CSP first and then relax only the lowest-enthalpy structures with SSCHA, e.g., within the 100~meV/atom threshold above the ground state \cite{Gao2025,Gao2026prb}. However, this risks missing phases that are high in enthalpy at 0~K but stabilized by anharmonicity at finite temperatures (e.g., bcc-Ti and bcc-Zr have 0~K enthalpies more than 100 meV/atom higher than their corresponding ground states but are stabilized at temperatures higher than 1000~K). 
Some high-pressure superconducting hydrides are synthesized by laser heating to temperatures near 2000~K \cite{Semenok2021}. 
Under these pressure-temperature conditions, the relevant free-energy landscape may differ qualitatively from a static 0 K enthalpy landscape. 
A phase formed during heating may be retained metastably after quenching or may transform during subsequent cooling or decompression. 
The high-temperature anharmonic landscape exploration being combined with kinetic considerations and simulated diffraction patterns may valuably help to identify and assign hydride phases observed experimentally, particularly when the experimentally recovered phase is not necessarily the thermodynamic ground state at 0 K.

\begin{figure*}
    \centering
    \includegraphics[width=1\linewidth]{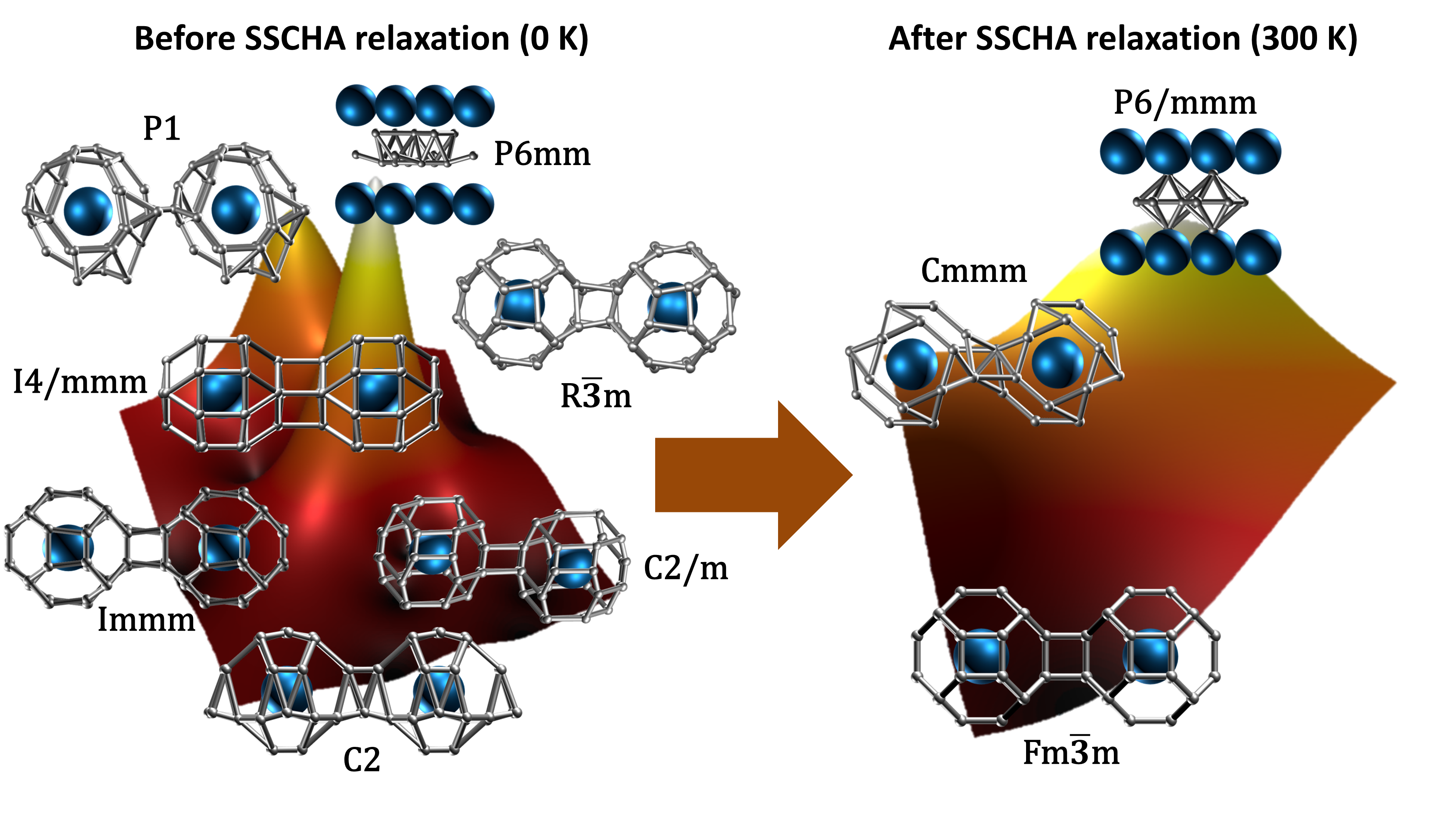}
    \caption{A comparison of energy landscapes before and after the SSCHA relaxation of structures (using local AL-MLIPs) preliminarily relaxed at 0~K (using VASP) showing the reduction in the number of LaH$_{10}$ polymorphs at 150~GPa when quantum anharmonicity is considered. The quantum anharmonic free-energy landscape at 300~K exhibits a simpler configuration space than the classical 0~K landscape. Both landscapes were generated using the dimensionality reduction method for fingerprint space implemented in STMng \cite{Valle2005}. The approach to constructing energy landscapes takes into account only energy minima and shows a smooth interpolation between all minima. It is clear that accounting for quantum motion leads to a smoother surface due to the disappearance of many minima highlighting the critical role of anharmonic effects in narrowing the search space during structural relaxation and ranking. }    
    \label{fig:fig_lah10_structures_and_energy_landscapes}
\end{figure*}

A structure search on the full quantum anharmonic free energy landscape can also benefit from its simplification compared to the classical 0~K landscape. 
Classical CSP at 150~GPa produces over twice as many polymorphs for LaH$_{10}$ and fails to identify $\mathrm{Fm}\bar{3}\mathrm{m}$ as the most stable, contradicting experiment \cite{Drozdov2019}. Figure~\ref{fig:fig_lah10_structures_and_energy_landscapes} illustrates this simplification, highlighting the practical advantage of anharmonic CSP in reducing candidate structures for costly property calculations.

A key limitation of SSCHA is its inability to handle tunneling and rotational modes in molecular crystals. This should be considered when applying our strategies to such systems. Overcoming this will require coupling the CSP with path-integral methods and modifying structure representations within CSP algorithms to account for non-local ion densities.

In summary, we have developed and validated three CSP approaches that incorporate quantum anharmonic effects through SSCHA accelerated by MLIPs and further parametrized by effective potentials. Using LaH$_{10}$ at 150~GPa, we show that:

\begin{enumerate}
    \item \textbf{Lightweight local AL-MLIPs trained from scratch} enable accurate CSP on the anharmonic free-energy landscape when combined with corrections from thermodynamic perturbation theory. Reusing lightweight local AL-MLIPs during SSCHA relaxation within CSP reduces the DFT cost by several times, but also lowers the accuracy of structural relaxation and ranking because of accumulated MLIP errors.
    
    \item \textbf{Large uMLIPs or foundation models}, such as Mattersim-5m, offer a promising alternative. They nearly eliminate the need for system-specific training of AL-MLIPs for SSCHA-based CSP, while maintaining robust SSCHA relaxations and enabling \textit{ab initio} structure ranking through thermodynamic perturbation theory corrections. However, fine-tuning the foundation model may still be necessary to obtain structures relaxed closer to the \textit{ab initio} level of theory.
    
    \item \textbf{Effective potentials trained on SSCHA data} provide a fast and scalable solution for large unit cells, enabling CSP on quantum anharmonic free-energy landscapes with up to 88 atoms per cell at manageable cost.
    
    \item \textbf{Quantum anharmonicity simplifies the search space}, reducing the number of plausible polymorphs compared to classical 0~K landscapes.

\end{enumerate}

Our work presents a step forward for efficient, quantum-accurate CSP in strongly anharmonic systems and highlights current limitations on this way. 
Future work should focus on improving uMLIP accuracy at high pressures, developing robust and computationally efficient protocols for training effective potentials, and extending the framework to systems where tunneling and rotational effects are significant.

\section{Methods}
\label{sec:methods}

\subsection{AL-MLIP selection and accuracy.}
\label{sec:mlip_selection}
\textbf{AL-MLIP selection.}
We used moment tensor potentials (MTPs) as AL-MLIPs.
To select the optimal MTP level, we relaxed the rhombohedral $\mathrm{R}\bar{3}\mathrm{m}$ LaH$_{10}$ structure to the cubic $\mathrm{Fm}\bar{3}\mathrm{m}$ structure within the SSCHA at 160~GPa and 300~K using local AL-MTPs of different levels.
We found that, for AL-MTPs of level eight, the free-energy error relative to the reference value obtained from SSCHA+DFT calculations is about 10~meV/atom before applying thermodynamic perturbation theory corrections and less than 1~meV/atom after applying them.

\textbf{Accuracy of AL-MLIPs.}
The training (validation) root-mean-squared errors (RMSEs) of the local AL-MTPs trained during the structure search were $9-32$ ($5-22$)~meV/atom for energies, $370-658$ ($322-478$)~meV/$\AA$ for forces, and $8.5-15.7$ ($2.9-6.0$)~GPa for stresses.
For reusable AL-MTPs, the training (validation) RMSE values were $10-105$ ($6-158$)~meV/atom for energies, $393-2122$ ($336-1000$)~meV/$\AA$ for forces, and $9.5-87.5$ ($3.5-99.3$)~GPa for stresses.
Validation RMSE values were estimated using the configurations selected for the calculation of thermodynamic perturbation theory corrections.
Detailed information about the MLIP RMSE values for each structure predicted on the anharmonic free-energy landscape with local AL-MTPs, reusable AL-MTPs, and the uMLIP Mattersim-5m can be found in Tables~S1--S5 in the Supplementary Materials.

For forces, the RMSE was computed as the square root of the mean squared difference between the MLIP-predicted and reference (DFT) force vectors, averaged over all atoms in the dataset: 
\begin{equation}
\text{RMSE}_{\text{forces}} = \sqrt{\frac{1}{N_{\text{atoms}}} \sum_{i=1}^{N_{\text{configs}}} \sum_{j=1}^{N_{\text{atoms}}^{(i)}} \left\| \mathbf{F}_{ij}^{\text{MLIP}} - \mathbf{F}_{ij}^{\text{DFT}} \right\|_2^2},
\end{equation}
where $\mathbf{F}_{ij}$ represents the force vector of atom $j$ in configuration $i$, $\left\| \cdot \right\|_2$ denotes the Euclidean norm defined as $\left\| \mathbf{F} \right\|_2 = \sqrt{F_x^2 + F_y^2 + F_z^2}$, and $N_{\text{atoms}} = \sum_i N_{\text{atoms}}^{(i)}$ is the total number of atoms across all configurations.

For stresses (virial tensors), the RMSE was calculated using the Frobenius norm of the $3 \times 3$ stress-tensor difference, averaged over the number of configurations:
\begin{equation}
\text{RMSE}_{\text{stresses}} = \sqrt{\frac{1}{N_{\text{configs}}} \sum_{i=1}^{N_{\text{configs}}} \left\| \boldsymbol{\sigma}_{i}^{\text{MLIP}} - \boldsymbol{\sigma}_{i}^{\text{DFT}} \right\|_F^2},
\end{equation}
where $\left\| \cdot \right\|_F$ is the Frobenius norm defined as:
\begin{equation}
\left\| \mathbf{A} \right\|_F = \sqrt{\sum_{p=1}^{3} \sum_{q=1}^{3} A_{pq}^2},
\end{equation}
and $\boldsymbol{\sigma}_i$ is the $3 \times 3$ stress tensor for configuration $i$ (converted to GPa). This formulation accounts for all nine components of the stress tensor, including both diagonal ($\sigma_{xx}$, $\sigma_{yy}$, $\sigma_{zz}$) and off-diagonal ($\sigma_{xy}$, $\sigma_{xz}$, $\sigma_{yz}$) elements, providing a complete measure of the stress-prediction accuracy.

\subsection{Computational details}
\label{sec:calculation_details}
\textbf{SSCHA-based structure prediction.}
In all cases, fixed-composition searches were performed in unit cells containing 11 atoms.
The first generation consisted of 30 candidates, while each subsequent generation consisted of 20 structures.
In the first generation, structures were created only using the random-symmetry operator.
Starting from the second generation, the soft-mode mutation and heredity operators were also used to generate structures.
The searches were stopped if the most stable structure did not change for 7 generations.

\textbf{DFT parameters.}
The DFT calculations used for MLIP training, thermodynamic perturbation theory corrections, and preliminary 0~K structural relaxations were performed within the generalized gradient approximation (GGA) for the exchange-correlation (XC) energy using the Perdew-Burke-Ernzerhof (PBE) parametrization \cite{Perdew1996a} and standard projector augmented-wave (PAW) \cite{Blochl1994} potentials for La and H, as implemented in the Vienna \textit{Ab initio} Simulation Package (VASP) \cite{Kresse1996}.
The kinetic-energy cutoff (500~eV) and $k$-point density ($0.05~2\pi/\AA$) were chosen on the basis of previous calculations by our group for this system \cite{Kruglov2020} and were additionally checked for convergence using the harmonic phonon dispersion curves of cubic LaH$_{10}$ with $\mathrm{Fm}\bar{3}\mathrm{m}$ symmetry.

\textbf{SSCHA calculations.}
Each SSCHA relaxation started from initial guesses for $\mathcal{R}$ and $\Phi$: $\mathcal{R}$ was obtained from a 0~K structural relaxation performed either with DFT or using a uMLIP, whereas $\Phi$ was obtained from harmonic force constants computed with a pretrained AL-MLIP or a uMLIP.
The upscaling SSCHA relaxation scheme outlined in Figure \ref{fig:sscha_mlip_common_scheme}, with $1\times1\times1$, $2\times2\times2$, and $4\times4\times4$ supercells for the first, second, and third stages, respectively, was used.
However, we found that the anharmonic free energy obtained at the end of the second and third stages converged to within $\sim$2~meV/atom for the LaH$_{10}$ structures selected for testing, indicating rapid convergence of the force constants with interatomic distance. For this reason, to speed up the calculations, we allowed USPEX to use the results from the second stage and to discard the third stage.
The numbers of configurations in the ensembles were 1000 and 3000 for the first and second stages of the SSCHA relaxation, respectively.
Each stage consisted of three substages: (1) relaxation of the centroid positions of the ions with a fixed cell; (2) variable-cell relaxation at fixed volume; and (3) variable-cell relaxation to the desired isotropic pressure, taking into account contributions arising from quantum fluctuations.
After the variable-cell relaxation at fixed volume, the pressure corrections were calculated as the mean deviation of the diagonal stress components over 50 configurations with the highest probability weights, selected from the ensemble with which the variable-cell SSCHA relaxation at fixed volume had converged.
The calculated pressure corrections were then subtracted from the target pressure.

The free-energy corrections up to second order within thermodynamic perturbation theory were calculated after the variable-cell relaxation to the corrected isotropic pressure using the formulation \cite{Landau2013}:
\begin{equation}
F_{\mathrm{DFT}} \simeq F_{\mathrm{}}+
\frac{1}{N_{\mathrm{at}}}\left[
\langle\Delta U\rangle
-\frac{1}{2k_{\mathrm B}T}
\left(U_{\mathrm{DFT}}-U_{\mathrm{}}-
\langle\Delta U\rangle\right)^2
\right],
\label{eq:thermodynamic_perturbation_corrections}
\end{equation}
given
\begin{equation}
\Delta U = U_{\mathrm{DFT}}-U_{\mathrm{}},
\label{eq:delta_U}
\end{equation}
where $F_{\mathrm{}}$ is the Helmholtz free energy obtained after the SSCHA relaxation with MLIP, $U_{}$ and $U_{DFT}$ are the total energies of configurations calculated using the MLIP and DFT, respectively; $k_b$ and $T$ are the Boltzmann constant and absolute temperature; and angle brackets denote averages over the configurations.
To calculate the free-energy corrections, 100 configurations with the highest probability weights were selected from the final ensemble with which the SSCHA relaxation to the corrected isotropic pressure had converged.
At all stages and substages, the SSCHA relaxation was considered converged when the meaningful factor became smaller than $0.001$ (for an explanation of this quantity, see \cite{Monacelli2021}).
The Gibbs free energy was obtained by adding the $PV$ term to the corrected Helmholtz free energy, where $P$ is the uncorrected isotropic pressure.

For the $\mathrm{Fm}\bar{3}\mathrm{m}$, $\mathrm{Cmmm}$, and $\mathrm{P}6/\mathrm{mmm}$ LaH$_{10}$ structures,
we performed additional SSCHA relaxations and Hessian calculations using actively learned level-20 MTPs to obtain accurate phonon dispersions and to assess dynamical stability at both the harmonic and anharmonic levels.
The harmonic phonon dispersion was calculated using the small-displacement method \cite{Ho1984}, as implemented in the ASE package \cite{Larsen2017}, with displacements of 0.05~\AA. The anharmonic phonon dispersion was obtained from the Hessian of the free energy within the SSCHA using an ensemble of 100,000 configurations. Higher-order quartic corrections were found to be negligible and were not included in the Hessian calculation. In both the harmonic and anharmonic cases, the calculations were performed using $2\times2\times2$ supercells.

\subsection{Training and Using TDEPs}
\label{sec:effective_potentials}

\textbf{Training TDEPs.}
For training the TDEPs, we collected data points only from those SSCHA ensembles in which more than 95\% of configurations have minimum interatomic distances no less than 2.0 $\AA$, 1.0 $\AA$, and 0.6 $\AA$ for La-La, La-H, and H-H atom pairs, respectively. 
The obtained dataset, which contained 14671 data points, was randomly divided into training (80\%) and validation (20\%) datasets, with 9039 and 2259 configurations, respectively.
These datasets were used for training TDEP-DPA3.
We trained four TDEP-DPA3 models from scratch with 2.4M parameters initialized using different seeds to obtain uncertainty estimates for relaxed structures. 

We also trained 11 TDEP-MTPs of different levels.
For training TDEP-MTP, we used the MaxVol algorithm \cite{Olshevsky2010} for selecting training configurations from the whole dataset.  
The numbers of training configurations for TDEP-MTPs of levels 8, 10, 12, 14, 16, 18, 20, 22, 24, 26, and 28 were 68, 114, 167, 229, 292, 387, 485, 589, 712, 827, and 949, respectively. 

\textbf{Accuracy of TDEPs.}
The RMSEs for energies of TDEP-DPA3 ranged from 0.8 to 3.8~meV/atom on the training dataset and from 0.3 to 1.5~meV/atom on the validation dataset.
The training RMSEs for energies of TDEP-MTPs of levels 8, 10, 12, 14, 16, 18, 20, 22, 24, 26, and 28 were 179, 130, 108, 118, 117, 97, 85, 77, 70, 65, and 70 meV/atom, respectively. 

\textbf{Using TDEPs within the CSP.}
The structure relaxations using TDEP within the CSP were performed using the LBFGS algorithm \cite{Liu1989} with FrechetCellFilter as implemented in ASE \cite{Larsen2017}.  
To control the ability of the TDEPs to correctly describe the relaxed structures, we used the extrapolation grade \cite{Podryabinkin2023} for TDEP-MTP and the ensemble dispersion estimate \cite{Carrete2023} of the energy for TDEP-DPA3.
At each relaxation step, the energies, forces, and stresses from four TDEP-DPA3 trained with different seeds were averaged.
The relaxed structure was considered correctly described by TDEP if its extrapolation grade was less than 10 (in the case of using TDEP-MTP) or if its energy dispersion was not higher than ~6 meV/atom (in the case of using TDEP-DPA3).

%%%%%%%%%%%%%%%%%%%%%%%%%%%%%%%%%%%%%%%%%55
\section{Acknowledgments}
% This work was supported by the Russian Science Foundation (Grant No. 19-72-30043).
Calculations with MTPs were performed on the Oleg cluster at the Skolkovo Institute of Science and Technology.
Calculations with uMLIPs, including Mattersim-5m, training and using TDEP-DPA3 were performed mainly on the Zhores cluster at the Skolkovo Institute of Science and Technology and partly on the computational resources of the HPC SEC shared research facilities at UNN. 

\section{Funding}
This work was supported by the Russian Science Foundation (Grant No. 19-72-30043).

\section{Author contributions}
Daniil Poletaev initiated and designed the project, performed the calculations, analyzed the data, and wrote the manuscript. Artem R. Oganov gave the idea and supervised the project, participated in discussions of the methods and results, and revised the manuscript.

\section{Data availability}
All data supporting the findings of this study are available within the paper, its Supplementary Information, and on the GitHub page: \url{https://github.com/Kurufinve/uspex-sscha-mlip-publication}.
Other additional data generated in this study are available from the authors upon reasonable request.

\section{Competing interests}
The authors declare no competing interests.

\bibliographystyle{elsarticle-num} 
\bibliography{library.bib}

\listoffigures
\listoftables

%%%%%%%%%%%%%%%%%%%%%%%%%%%%%%%%%%%%%%%%%

\end{document}